\documentclass[conference,9pt]{IEEEtran}
\IEEEoverridecommandlockouts

\usepackage{cite}
\usepackage{amsmath,amssymb,amsfonts}
\usepackage{algorithmic}
\usepackage{algorithm}
\usepackage{graphicx}
\usepackage{textcomp}
\usepackage{xcolor}
\def\BibTeX{{\rm B\kern-.05em{\sc i\kern-.025em b}\kern-.08em
    T\kern-.1667em\lower.7ex\hbox{E}\kern-.125emX}}
\usepackage{stfloats}

\begin{document}

\title{Escaping Barren Plateau: Co-Exploration of Quantum Circuit Parameters and Architectures
}

\author{
\IEEEauthorblockN{Yipei Liu\textsuperscript{\dag}, Yuhong Song\textsuperscript{\dag}, Jinyang Li\textsuperscript{\dag}, Qiang Guan\textsuperscript{\ddag}, Cheng-chang Lu\textsuperscript{*}, Youzuo Lin\textsuperscript{\S}, Weiwen Jiang\textsuperscript{\dag}}
\IEEEauthorblockA{\textsuperscript{\dag}George Mason University, 
\textsuperscript{\ddag}Kent State University, 
\textsuperscript{*}Qradle Inc., 
\textsuperscript{\S}University of North Carolina at Chapel Hill}

}


\maketitle

\begin{abstract}
Barren plateaus (BP), characterized by exponentially vanishing gradients that hinder the training of variational quantum circuits (VQC), present a pervasive and critical challenge in applying variational quantum algorithms to real-world applications. It is widely recognized that the BP problem becomes more pronounced with an increase in the number of parameters. This work demonstrates that the BP problem manifests at different scales depending on the specific application, highlighting the absence of a universal VQC ansatz capable of resolving the BP issue across all applications. Consequently, there is an imminent need for an automated tool to design and optimize VQC architectures tailored to specific applications. To close the gap, this paper takes Variational Quantum Eigensolvers (VQEs) as a vehicle, and we propose a novel quantum circuit parameter and architecture co-exploration framework, namely AntiBP. Experimental results demonstrate that AntiBP effectively avoids the BP issue for circuits that are not under-parameterized in noise-free environments. Furthermore, AntiBP significantly outperforms baseline VQEs in noisy environments.
\end{abstract}


\section{Introduction}

The Variational Quantum Circuit (VQC) framework has emerged as a cornerstone for quantum computing applications in the Noisy Intermediate-Scale Quantum (NISQ) era. By leveraging classical optimization to train quantum circuits, VQCs provide a promising pathway for solving computational problems that are challenging for classical methods, such as computational chemistry \cite{motta2022emerging,feniou2023overlap,cao2019quantum,izmaylov2019unitary,izsak2023quantum} and material science \cite{bauer2020quantum,alexeev2024quantum, teale2022dft,reiser2022graph,westermayr2021perspective}, optimization and machine learning \cite{jiang2021co, wang2021exploration, jiang2021machine, stein2022quclassi, jing2022rgb, zeng2022multi, cerezo2022challenges}. Algorithms like Variational Quantum Eigensolvers~(VQE) \cite{Peruzzo2014}, Quantum Approximate Optimization Algorithm (QAOA) \cite{a12020034}, and quantum-enhanced machine learning exemplify the utility of VQCs in tackling real-world problems within the constraints of current quantum hardware. However, a common and critical challenge faced by all these algorithms is the barren plateau (BP) issue, where exponentially vanishing gradients \cite{Arrasmith_2022} severely hinder the optimization and scalability of VQCs, especially as the size and complexity of quantum circuits grow. 
This problem hampers optimization efficiency and limits the expressiveness and convergence of VQC circuits, especially on NISQ devices.

One prominent approach to mitigating the BP problem focuses on careful parameter initialization, as random initialization often results in flat loss landscapes, as demonstrated by \cite{mcclean2018barren}. Recent works, such as \cite{kulshrestha2022beinit}, have developed algorithms for improved initialization to avoid the BP problem. However, these methods are computationally costly and, more importantly, fail to enhance performance in noisy environments \cite{Limitations2021}. Beyond initialization, the architecture of the variational ansatz also plays a significant role in addressing the BP problem. For example, \cite{grant2019initialization} proposes appending an inverse circuit to the original VQC to alleviate BP issues. While this approach mitigates BP, it doubles the parameter count, significantly degrading performance in noisy environments. Notably, all these studies rely on a backbone VQC architecture, within which the BP problem inherently persists.

\begin{figure}[t]
\centerline{\includegraphics[width=1.0\linewidth]{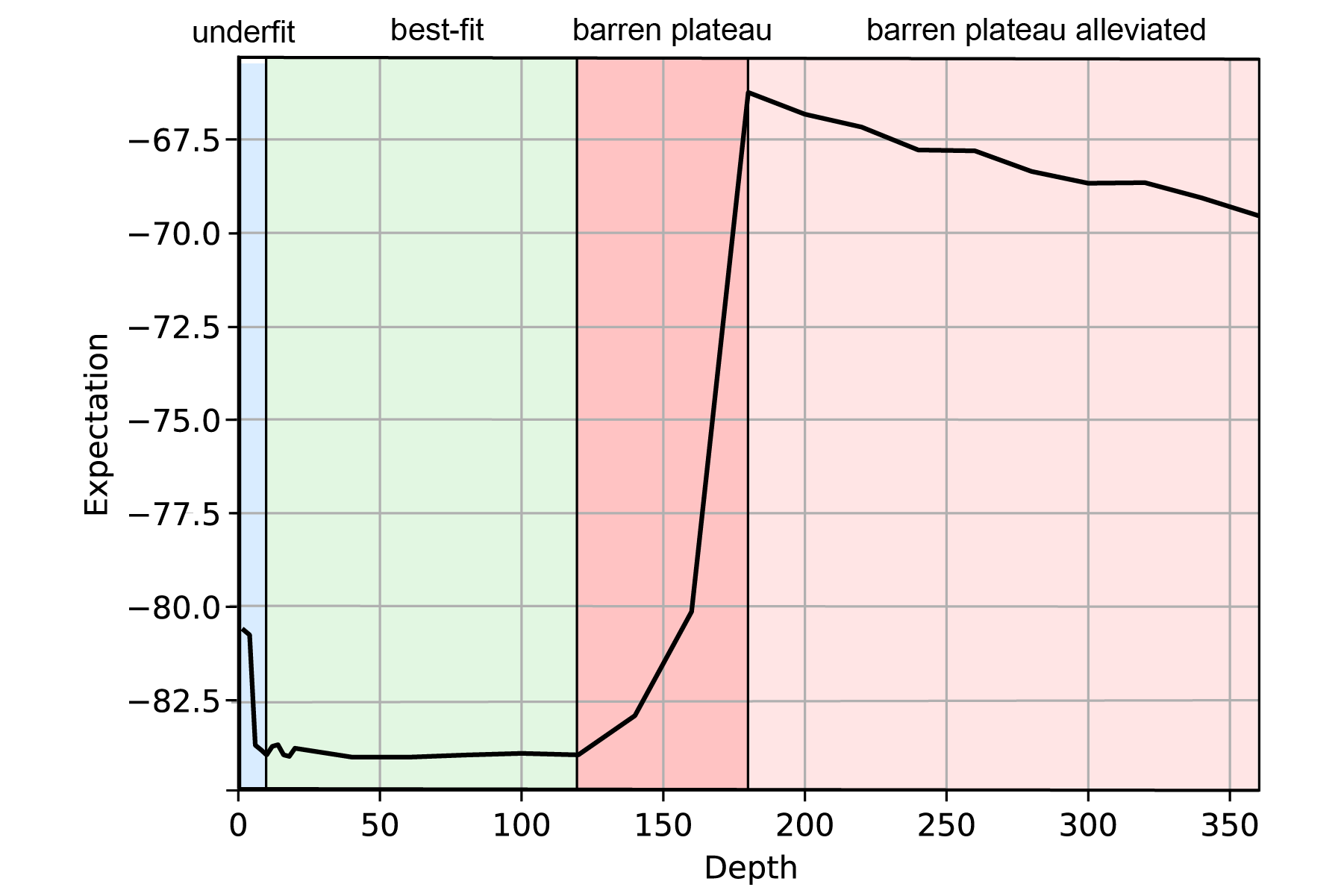}}
\caption{Performance of VQE on $H_2O$ molecule with different circuit depths, revealing four performance regions in the VQE optimization process}     
\label{fig:ori}
\end{figure}

In this work, we rethink the BP problem, aiming to identify a backbone architecture for a given application where the BP issue inherently does not persist. To enable this, we use the VQE as a vehicle and analyze the optimization performance of backbone quantum circuit architectures as the circuit depth increases, which is proportional to the number of parameters \cite{Larocca2022}. Results, as shown in Figure \ref{fig:ori}, reveal four distinct performance regions based on circuit depth: (1) Underfit Region, where a lack of parameters limits the VQE's ability to predict ground state energy accurately; (2) best-fit Region, representing the optimal VQE architecture for the application; (3) Barren Plateau Region, where the BP issue emerges, leading to optimization stagnation; and (4) BP Alleviated Region, where performance gradually improves due to overparameterization despite the associated inefficiencies.

This motivated us to explore an alternative approach --- instead of relying on complex optimization techniques for initialization, it is possible to identify a robust backbone architecture that inherently falls within the best-fit region, thereby avoiding the BP problem.
Now, the key is to identify the best-fit region, particularly if we can find a uniform best-fit region for different applications. Unfortunately, our analysis reveals that the best-fit regions vary significantly between applications; for instance, the VQE for $H_2O$ exhibits a distinct best-fit region compared to that for $LiH$. Furthermore, in noisy environments, the best-fit region contracts sharply, making the search process increasingly challenging.

In this work, we propose an automated framework, namely \textit{AntiBP}, to identify the VQE architecture that falls within the best-fit region by co-exploring quantum circuit parameters and architectures. The objective is to ensure that the optimizer can consistently converge to a solution (i.e., quantum circuit architecture and parameters) for a given deep enough quantum circuit.
To achieve this goal, a dynamic circuit adjustment algorithm is proposed that automatically determines which quantum gates should be retained during the optimization process. By evaluating each gate's contribution to the overall objective, this method preserves the expressive power of the circuit while optimizing its structure to avoid barren plateaus.

The main contributions of this paper are as follows:

\vspace{-\topsep}
\setlength{\parskip}{5pt}
\setlength{\itemsep}{0pt plus 1pt}
\begin{itemize}
    \item To the best of our knowledge, this is the first pilot study to address the barren plateaus problem by co-exploring the variational quantum circuit architecture and parameters.
    \item We developed an efficient search framework, namely AntiBP, to optimize both parameters in a unified optimization loop for a given application, which also shows the capability to identify high-fidelity quantum circuits in a noisy environment.

    \item We carried out a set of experiments on standard quantum chemistry problems to demonstrate the effectiveness of AntiBP. The proposed method not only improves convergence rates but also enhances the accuracy of the ground-state energy estimations, offering a promising direction for scaling VQEs on NISQ devices. 

\end{itemize}
\vspace{-\topsep}

\section{Related work}

The barren plateau (BP) problem, characterized by exponentially vanishing gradients in VQC training, poses a significant obstacle to the scalability and efficiency of quantum machine learning and optimization tasks. 

This issue has been extensively studied, with solutions ranging from initializations to circuit design strategies.

\begin{figure}[t]
\centerline{\includegraphics[width=0.5\textwidth]{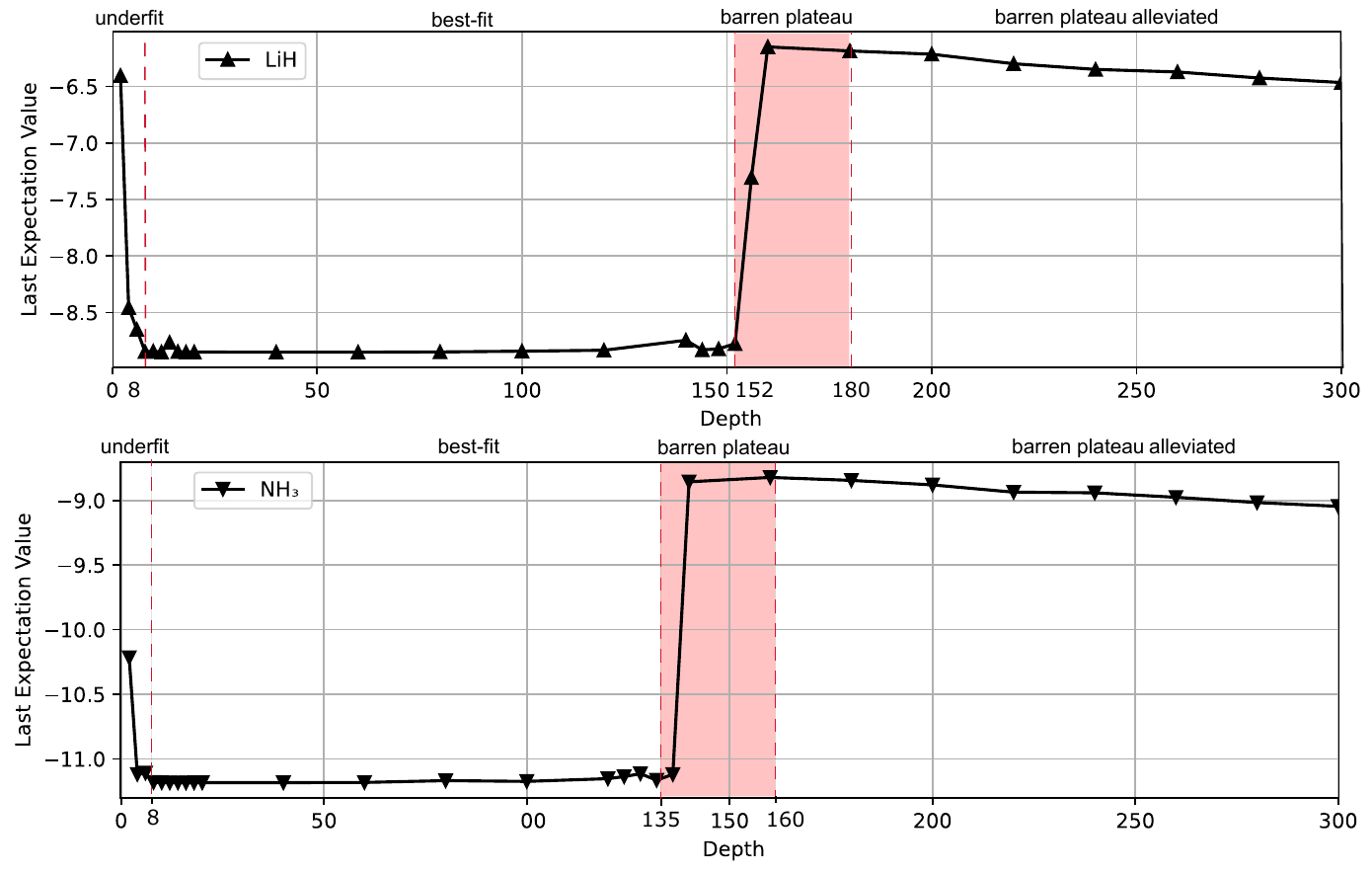}}
\caption{Performance regions shift along with different applications, taking VQE optimization for $LiH$ and $NH_3$ molecular as an example.}     
\label{fig:NH3}
\end{figure}

One prominent approach to mitigating the BP problem \cite{Qi2023} focuses on careful parameter initialization. Work \cite{mcclean2018barren} first identifies the BP problem, demonstrating that random initialization often leads to flat loss landscapes in deep quantum circuits. \cite{grant2019initialization} addresses this by proposing the ``IdentityBlock" method, which appends an inverse circuit to the original VQC to create an identity circuit initialized with random parameters. This strategy effectively doubles the parameter volume, aiming to preserve high gradient magnitudes during the first optimization epoch and transition the model into the BP alleviation stage.
However, the observed performance gains from this method appear highly dependent on the specific application and circuit architecture. 
Our study shows that doubling parameters may still result in performance being trapped within the BP range.

Efforts to mitigate the BP problem have also explored variable structure ansatzes to design shallow and trainable VQCs. Some approaches \cite{Bilkis2023} build circuits from scratch, aiming to avoid BP by reducing depth. However, such methods often overlook application-specific requirements, as seen in VQEs \cite{PRXQuantum.2.020329}, where domain-specific features are crucial. As a result, circuits optimized purely for BP mitigation may lack the necessary characteristics for effective problem-solving.
Other methods \cite{Du2022} use gate sampling to construct circuits, but this introduces significant randomness and high computational costs. As circuit depth increases, the exponentially growing configuration space makes sampling inefficient and impractical for identifying optimal designs. 
Some approaches \cite{Sim_2021} improve optimization efficiency by iteratively fine-tuning and pruning circuit structures. However, removing critical gates early to escape the BP may limit the circuit's expressiveness and problem-solving ability.

The above limitations highlight the need for an automated framework capable of adapting circuit designs to different quantum applications, ensuring both effective BP mitigation and problem-specific optimization.

\section{Observation and Motivation}

Figure \ref{fig:ori} shows four optimization regions by circuit depth. This section further demonstrates that the best-fit region is application-dependent and changes significantly when experiments run on noisy devices.

\subsection{Observation 1: Regions Shift with Different Applications} 

To investigate the correlation between the optimization region and applications, we carry out a set of experiments on different molecules.
To eliminate the potential impact of the different quantum circuit scales, we normalized the molecular structures by
constraining the number of electrons to 6 and the number of spatial orbitals to 6.
As such, each VQE has 12 qubits.
Figure~\ref{fig:NH3} reports the results. These figures clearly show that the barren plateau region of the energy estimation VQE for molecules, $LiH$ and $NH_3$, are quite different.

This observation indicates that the optimization regions (i.e., circuit depth) are different in terms of applications.
This brings high challenges in designing a VQE circuit for an unknown problem: we cannot determine the best-fit region to avoid the barren plateau in advance.
One straightforward solution is to manually find the best-fit region through a series of trial and error, which is computationally costly and time-consuming.

\textbf{Motivation 1:} This highlights the need for an automated VQE architecture search that can adjust the circuit structure based on the given application during optimization.

\begin{figure}[t]
\centerline{\includegraphics[width=0.5\textwidth]{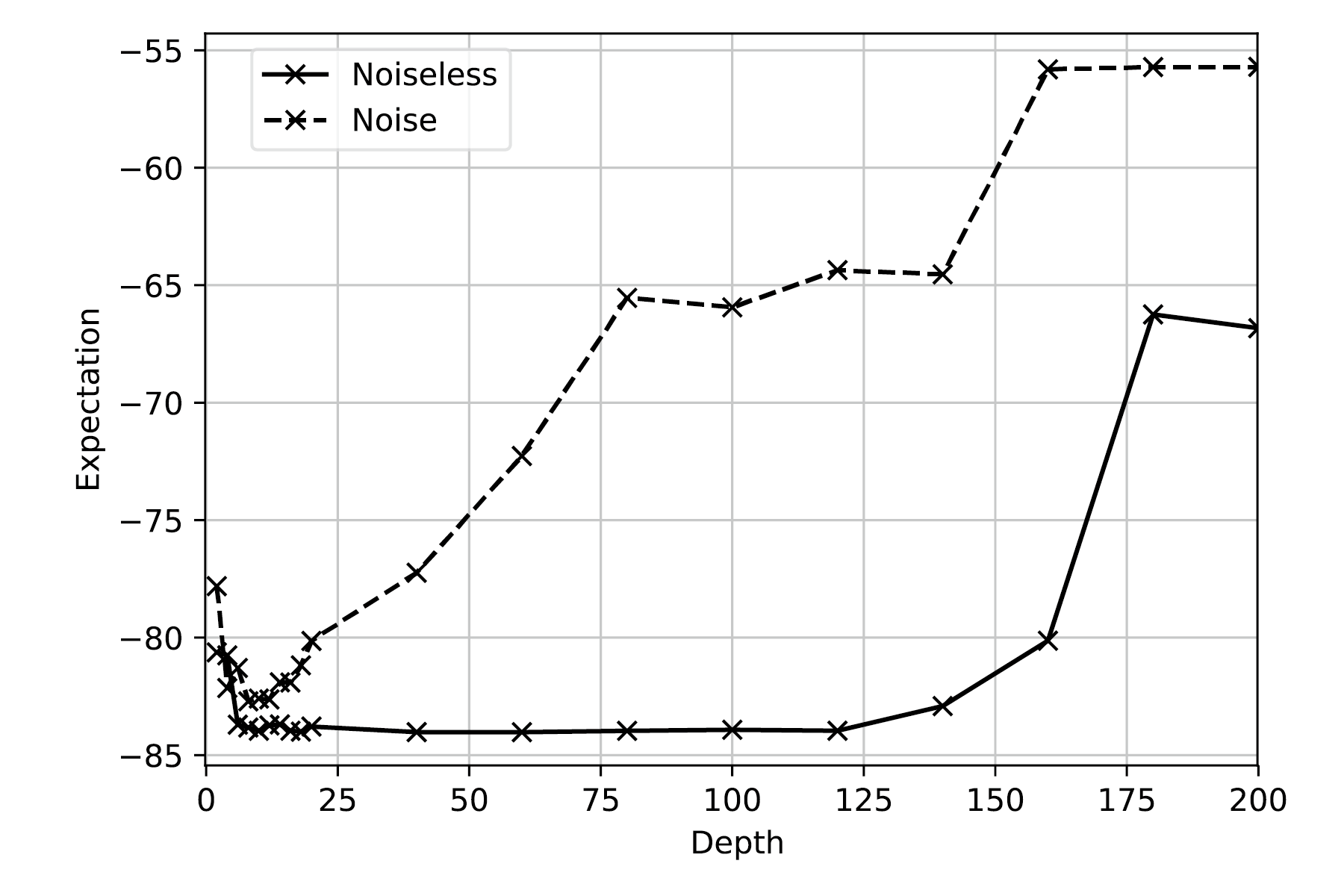}}
\caption{Performance of VQE with varying circuit depths for $H_2O$ molecular under noise-free and noisy environments.}     
\label{fig:noiseIden}
\end{figure}

\begin{figure*}[b]
\vspace{20pt}
\centerline{\includegraphics[width=0.75\linewidth]{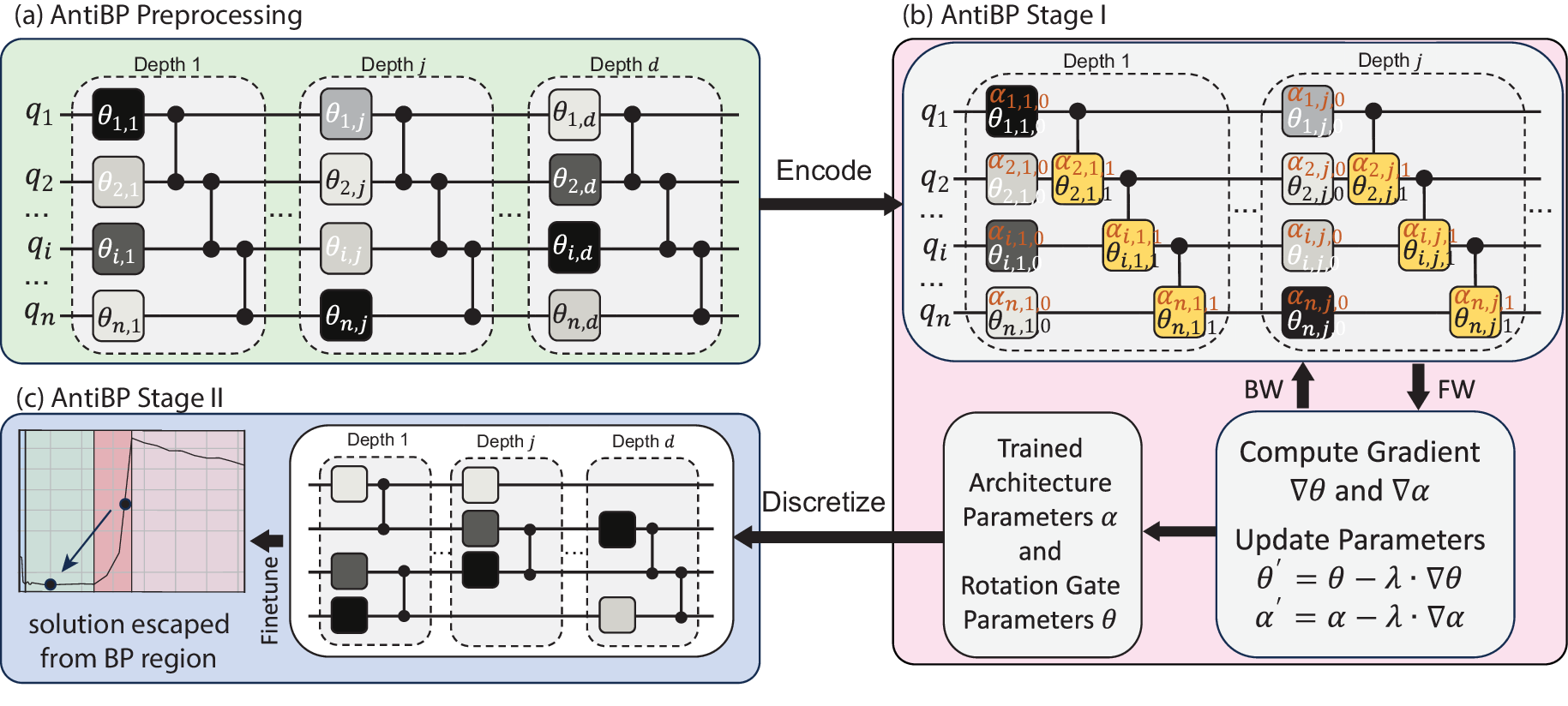}}
\vspace{-8pt}
\caption{Overview of the proposed AntiBP framework: (a) AntiBP preprocessing: for encoding VQE with gate parameters only to VQE with both gate parameters and architecture parameters; (b) AntiBP Stage 1: Co-optimize gate and architecture parameters; (c) AntiBP Stage 2: Finetuning gate parameters with fixed architecture parameters}
\label{fig:frame}
\end{figure*}

\subsection{Observation 2: Noise on devices making the search process more challenging}  

All the above noise-free results demonstrate a related wide best-fit region, while the high-level noise on today's quantum device can greatly shrink the best-fit region, exacerbating the challenge for VQE architecture search.
To showcase the impact of noise on the optimization region, we apply the same VQE for results in Figure \ref{fig:ori} and execute it on noisy devices.
Figure \ref{fig:noiseIden} reports the results.
We observe that even in the best-fit region (identified by noiseless experiments), the performance of VQE is degraded sharply, along with the increase in circuit depth.
Moreover, the best region for the optimization under noise is shapely shrunk, which appears around the depth of 12.

\textbf{Motivation 2:} This observation further emphasizes the critical need to automatically identify the backbone quantum circuit architectures in order to eliminate the effects of the barren plateau and make the VQE robust to quantum noise.

\section{AntiBP Framework}  

Inspired by the above observations, we propose a novel VQE architecture parameter and gate parameter co-exploration framework, namely AntiBP, which includes two stages of optimizations: (1) Circuit architecture search --- During optimization, we need to dynamically adjust the circuit by adding or removing gates based on their impact on the loss function. (2) Fine-tuning --- Once the VQE architecture is determined, the VQE circuit with a determined architecture will be applied to optimize the gate parameters only to ensure it reaches optimal performance for VQE tasks.

\subsection{Overview}

Figure \ref{fig:frame} shows the overview of the proposed AntiBP framework.
It contains three major processes: (1) Preprocessing, encoding a given input VQE circuit with only gate parameters; (2) Stage 1, co-optimizing the gate parameters and architecture parameters in a holistic optimization loop; and (3) Stage 2, fine-tuning gate parameters with identified architecture parameters in stage 1.

In the following, we will introduce the preprocessing and stage 1, respectively.
Once the architecture parameters have been determined, stage 2 can be conducted with hybrid classical-quantum optimization as other VQE optimizations.

\subsection{AntiBP Preprocessing}  
The conventional VQE only contains the gate parameters $\theta$, as shown in Figure \ref{fig:frame}(a).
One key step to enable AntiBP is to associate a set of additional parameters to the original quantum gates, such that both type of parameters can be optimized in a holistic optimization loop.

In this framework, every parameterized gate \( G(\theta) \) in \( \mathcal{C} \), where \( \theta \) represents the gate's original parameter. In our approach, the actual operation becomes \( G(\alpha \cdot \theta) \), where \( \alpha \) is a control parameter that causes the gate to have no effect when \( \alpha = 0 \) or to be executed normally with parameter \( \theta \) when \( \alpha = 1 \). Therefore, when exploring the circuit architecture, \( \alpha \) can take values of 0 or 1 to determine whether each gate is active or inactive, guiding our pruning strategy.

For controlled gates that are not initially parameterized, such as the CZ, CX, or CY gates, we redefine them as parameterized gates to increase flexibility. Specifically, we replace these gates with controlled rotations, such as \( \text{CRZ}(\pi \cdot\beta) \), \( \text{CRX}(\pi \cdot\beta) \), or \( \text{CRY}(\pi \cdot\beta) \), where \( \beta \in \{0, 1\} \) is a tunable parameter. This adjustment allows us to control the gate’s effect: when \( \beta = 0 \), the gate has no effect, effectively pruning this operation; when \( \beta = 1 \), the gate performs its original function (e.g., acting as a CZ, CX, or CY gate).

Mathematically, the original circuit \( \mathcal{C} \) consisted of parameterized gates with parameters \( \theta \) and controlled gates like CZ, CX, and CY without tunable parameters. In our modified framework, the circuit \( \mathcal{C}(\vec{\theta}, \vec{\alpha}, \vec{\beta}) \) is expressed as: \[ \mathcal{C}(\vec{\theta}, \vec{\alpha}, \vec{\beta}) = \prod_{i} G_i(\alpha_i \cdot \theta_i) \prod_{j} \text{CR}(\pi \cdot \beta_j), \] where \( G_i(\alpha_i \cdot \theta_i) \) represents each parameterized gate with an added control parameter \( \alpha_i \), determining its activation (with \( \alpha_i = 0 \) deactivating the gate and \( \alpha_i = 1 \) executing it with parameter \( \theta_i \)), and \( \text{CR}(\pi \cdot \beta_j) \) represents the controlled rotation gates with a control parameter \( \beta_j \) that activates the gate to function as the original CZ, CX, or CY gate when \( \beta_j = 1 \), or deactivates it when \( \beta_j = 0 \).

This modified circuit structure enables dynamic pruning and adjustment of the architecture during optimization. By setting appropriate values for the \( \alpha \) and \( \beta \) parameters, the circuit can explore different architectures by selectively activating or deactivating specific gates. This flexibility enables the model to achieve a balance between expressiveness and computational efficiency, therefore mitigating issues such as the barren plateau by avoiding redundant gates and focusing on the most impactful operations. By optimizing both \( \vec{\alpha} \) and \( \vec{\beta} \) throughout the optimization process, we can dynamically reconfigure the circuit to control each gate's presence and function, enabling efficient VQE convergence and simultaneous expressiveness and parameter optimization.

\begin{algorithm}[t]
\caption{AntiBP Architecture Search}
\label{alg:sig_arch_search}
\begin{algorithmic}[1]
	\STATE \textbf{Input:} Quantum circuit \( \mathcal{C}_{\text{input}} \)
	\STATE \textbf{Output:} Architecture-searched circuit \( \mathcal{C}_{\text{opt}} \)
	
	\STATE Initialize \( \vec{\theta} \) randomly, initialize \( \vec{\alpha} \gets 1 \), \( \vec{\beta} \gets 1 \)
	
	\STATE Construct architecture search circuit \(
	\mathcal{C}(\vec{\theta}, \vec{\alpha}, \vec{\beta})\):
	\STATE \hspace{1cm} Replace each gate \( G_i \) in \( \mathcal{C}_{\text{input}} \) with: \( G_i(\alpha_i \cdot \theta_i) \)
	\STATE \hspace{1cm} Replace each CZ gate in \( \mathcal{C}_{\text{input}} \) with: \( \text{CRZ}(\pi \cdot \beta_j) \)

	\WHILE{epoch}
	\STATE Compute expectation from circuit output $ |\psi(\vec{\theta})\rangle$:
	\[
	E(\vec{\theta}) = \langle \psi(\vec{\theta}) | H | \psi(\vec{\theta}) \rangle
	\]
	\STATE Update \( \vec{\theta}, \vec{\alpha}, \vec{\beta} \) via backpropagation:
	\[
	\vec{\theta}, \vec{\alpha}, \vec{\beta} \gets \text{Optimizer}(\vec{\theta}, \vec{\alpha}, \vec{\beta}, \nabla E)
	\]
	\ENDWHILE
	
	\STATE Prune gates based on final \( \vec{\alpha}, \vec{\beta} \):(0: prune, 1: keep)
	
	\RETURN Construct \( \mathcal{C}_{\text{opt}} \) with selected gates
\end{algorithmic}

\end{algorithm}

\subsection{AntiBP Stage 1}

To achieve a binary-like behavior for the architecture parameters \( \alpha \) and \( \beta \), we introduce a sigmoid function with a steep slope controlled by a parameter set to 50. This ensures that the output values of \( \text{sigmoid}(x) \) are nearly binary (close to 0 or 1) during VQE optimization, effectively forcing the architecture parameters \( \alpha \) and \( \beta \) to act as binary gates. These binary-like values allow the architecture search to dynamically determine whether a gate is active (\( \alpha_i, \beta_j \approx 1 \)) or pruned (\( \alpha_i, \beta_j \approx 0 \)).

During VQE optimization, the sigmoid function acts as a differentiable approximation of binary gating, allowing the use of backpropagation to compute gradients for \( \alpha \) and \( \beta \) through the chain rule. The loss function, defined as the expectation value:
\(
E(\vec{\theta}, \vec{\alpha}, \vec{\beta}) = \langle \psi(\vec{\theta}, \vec{\alpha}, \vec{\beta}) | H | \psi(\vec{\theta}, \vec{\alpha}, \vec{\beta}) \rangle
\)
depends on the output quantum state \( |\psi\rangle \), which in turn depends on \( \alpha \), \( \beta \), and \( \theta \). Employing a steep sigmoid ensures that the gradients with respect to \( \alpha \) and \( \beta \) remain significant, enabling efficient optimization using standard gradient descent methods.

Algorithm \ref{alg:sig_arch_search} outlines the proposed AntiBP algorithm, Sigmoid-Parameterized Architecture Search, which operates through the following key steps. First, an initial potential circuit \( \mathcal{C}(\vec{\theta}, \vec{\alpha}, \vec{\beta}) \) is generated, where gates are parameterized by control parameters \( \alpha \) and rotation parameters \( \beta \). These parameters allow flexibility in determining the gate configurations during the search process. Next, the architecture and parameter search is performed iteratively, where \( \alpha \) and \( \beta \) are adjusted using backpropagation to explore various configurations and identify the active gates required for an efficient architecture. Based on the search results, an optimized circuit \( \mathcal{C}_{\text{opt}} \) is constructed by pruning inactive gates. Finally, the optimized circuit \( \mathcal{C}_{\text{opt}} \) undergoes fine-tuning of its parameters \( \vec{\theta} \) to achieve optimal performance for the target application.

Our approach begins with a joint search over circuit architecture and parameters. Starting with the initial potential circuit \( \mathcal{C}(\vec{\theta}, \vec{\alpha}, \vec{\beta}) \), we iteratively adjust the control parameters \( \alpha \) and \( \beta \) to explore different configurations. During this process, \( \alpha \) values determine which parameterized gates \( G_i(\alpha_i \cdot \theta_i) \) are active, while \( \beta \) values selectively activate controlled rotation gates, such as CRZ, CRX, and CRY, by setting \( \beta_j = 1 \) for activation or \( \beta_j = 0 \) for deactivation. This search phase aims to identify a compact circuit configuration that maintains sufficient expressiveness while avoiding redundant gates, thus reducing the likelihood of encountering barren plateaus.
Through this iterative search, we obtain an optimized subset of gates and parameters. 
The result is a refined circuit with only the most effective gates, improving both computational efficiency and expressiveness.

With the relevant gates identified, we construct the optimized circuit \( \mathcal{C}_{\text{opt}} \), which includes only the active gates from the search phase. This new circuit structure is then fine-tuned to achieve optimal performance. In this phase, we keep the circuit architecture fixed and focus on refining the remaining parameters \( \vec{\theta} \) associated with the active gates. Fine-tuning enhances accuracy further by enabling the VQE to converge precisely on the target solution, leveraging a simplified and more efficient circuit architecture.

\section{Experimental Results}

We evaluate our framework on a quantum circuit-based task designed for benchmarking. Specifically, we use a 14-qubit quantum circuit to investigate the effect of depth on circuit performance. For each depth level, we randomly generate single-qubit gates (RX, RY, or RZ) for each qubit. To increase entanglement \cite{Somma2004}, we apply controlled-Z (CZ) gates between adjacent qubits in the circuit.

We analyze the performance of circuits as a function of depth, exploring how increasing the depth impacts circuit performance. Additionally, we compare the performance of three types of circuits at various depths: randomly generated circuits, circuits optimized through architecture search, and circuits constructed with identity blocks. This comparison allows us to evaluate the effectiveness of each approach under different depth conditions.

To evaluate the robustness of our framework under noisy conditions, we simulate quantum circuits in the presence of depolarizing noise. Depolarizing noise is a widely used quantum noise model that captures the impact of errors occurring in quantum operations. In this model, with a small probability \( p \), the quantum state is replaced by a mixed state, effectively modeling random errors. For example, a depolarizing noise with \( p = 0.001 \) means there is a 0.1\% chance of an error being applied to each operation in the circuit.

This type of noise reflects realistic imperfections encountered in quantum hardware, such as gate operation errors, decoherence, and environmental disturbances. By introducing depolarizing noise into the circuits, we analyze how such errors affect the performance of various circuit designs, including random circuits, circuits optimized through architecture search, and circuits constructed with identity blocks. This analysis provides a comprehensive understanding of the resilience of different circuit architectures in noisy environments.

\subsubsection{Effective Evaluation in Noiseless Environment}
\begin{figure}[t]
\centerline{\includegraphics[width=0.5\textwidth]{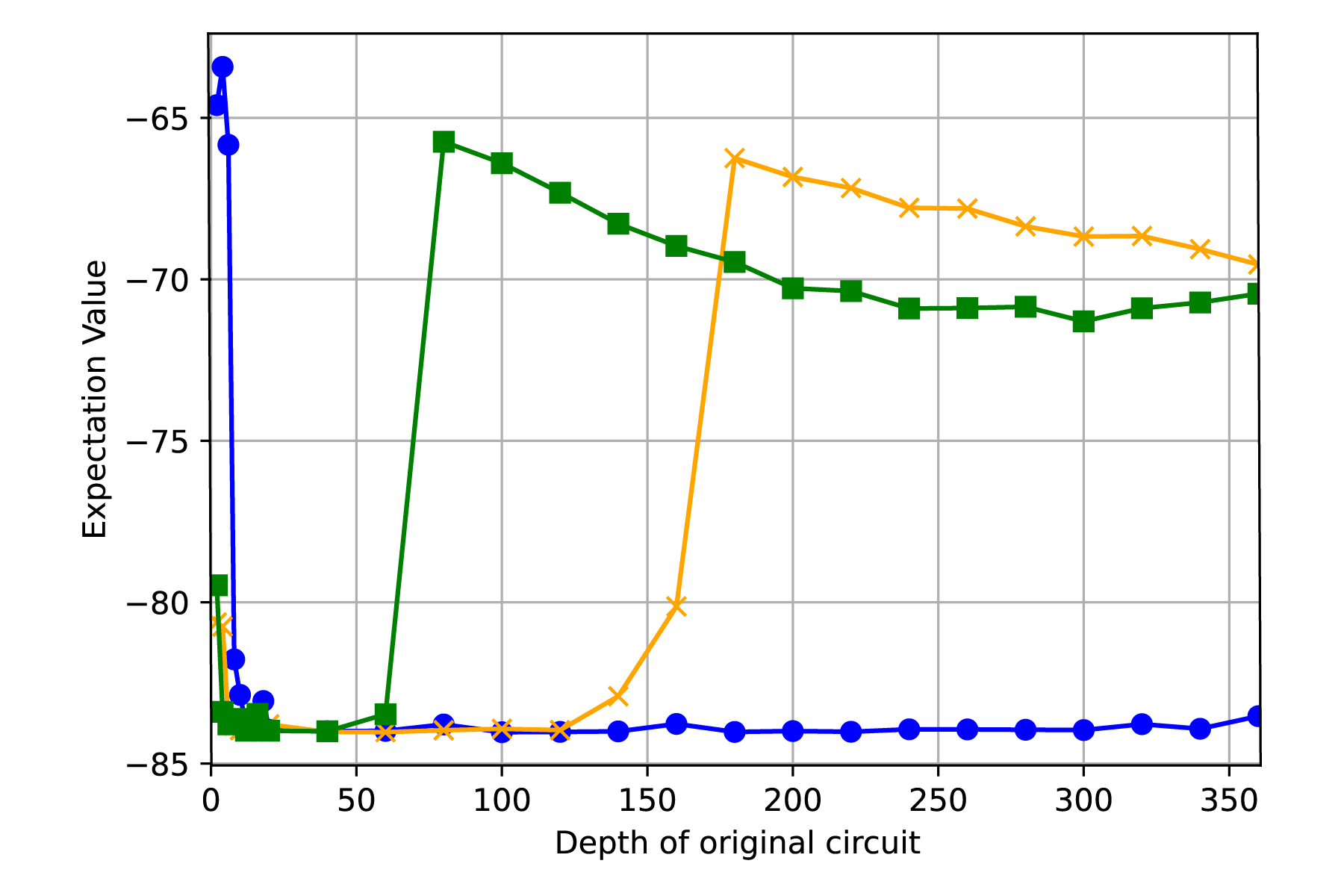}}
\caption{AntiBP consistently achieves high performance when the original circuit depth is larger than 50.}     
\label{fig:noiseless}
\end{figure}

Figure~\ref{fig:noiseless} reports the performance of different circuit designs in a noiseless environment as the circuit depth increases. The y-axis represents the expectation values, where lower values indicate better results. From the results, it is evident that our proposed architecture-optimized circuits outperform both identity block circuits and random circuits.

In the noiseless environment, the architecture-optimized circuits consistently achieve the lowest expectation values across all depths. This improvement can be attributed to the gate pruning process during architecture search, which identifies and removes gates that contribute to barren plateaus. By optimizing the circuit architecture, we significantly reduce the impact of redundant gates, allowing the circuit to maintain high performance regardless of depth. This is particularly evident as the depth increases, where the architecture-optimized circuits remain unaffected by the limitations of barren plateaus that typically hinder the optimization process in deeper circuits.

Compared to identity block circuits, the architecture-optimized circuits demonstrate superior performance and stability. Identity block circuits show better results than random circuits (baseline) as depth increases, particularly in mitigating the barren plateau phenomenon. This is evident from Figure~\ref{fig:noiseless}, where the identity block circuits (green) outperform the random circuits (orange) at deeper depths, demonstrating that the structured identity blocks provide some resilience to the barren plateau problem. However, this improvement is limited to alleviating the issue rather than fully solving it.

While identity block circuits achieve better results than the baseline, their performance plateaus and fails to converge to the globally optimal expectation values. This suggests that the static structure of the identity block design can only partially address the barren plateau problem, lacking the capability to adapt to deeper circuits. In contrast, the architecture-optimized circuits (blue), through dynamic pruning of barren plateau-causing gates, consistently achieve significantly better results and maintain robust performance regardless of depth. 
 This shows that the architecture search method alleviates barren plateaus and guides optimization toward the global optimum, which identity block circuits cannot reach.

These results validate the effectiveness of our architecture search process, which not only improves circuit expressiveness but also ensures the performance against the increased circuit depth. The ability to adaptively prune redundant gates ensures that the optimized circuit can efficiently achieve low expectation values without being constrained by depth.

\begin{table*}[!ht]
\caption{Comparison of number of gates and energy on $H_2O$ and $H_2$ molecular under noisy environment.}
\def\arraystretch{0.9}\begin{center}
\begin{tabular}{|c|c|c|c|c|c|c|c|c|c|}
\hline
\textbf{Molecule}& \textbf{Qubits}& \textbf{Depth} & \textbf{Method} & \textbf{\# 1-Q G} & \textbf{\# 2-Q G} & \textbf{Ref. Energy}& \textbf{Energy}& \textbf{Energy Gap} & \textbf{Improv.} \\
\hline
 &  &  & Vallina & 840 & 780 &  & -72.2699 & 11.7543  & baseline \\ 
 $H_2 O$ & 14 & 60 & IDBlock & 1680 & 1560 & -84.0242 & -54.9790 & 29.0451  & -14.71\% \\ 
 &  &  & AntiBP & 428 & 372 &  & -77.1156 & 6.9085  & 41.23\% \\ \hline
 &  &  & Vallina & 1120 & 1040 &  & -64.1854 & 19.8387  & baseline \\ 
$H_2 O$ & 14 & 80 & IDBlock & 2240 & 2080 & -84.0242 & -55.5399 & 28.4843  & -4.36\% \\ 
 &  &  & AntiBP & 867 & 660 & & -73.3062 & 10.7179  & 45.97\% \\ \hline
 &  &  & Vallina & 2240 & 2080 &  & -55.8122 & 28.2119  & baseline \\ 
$H_2 O$ & 14 & 160 & IDBlock & 4480 & 4160 & -84.0242 & -55.7004 & 28.3237  & 0.39\% \\ 
 &  &  & AntiBP & 1760 & 1711 &  & -64.1382 & 19.9960  & 29.51\% \\ \hline

 &  &  & Vallina & 240 & 180 &     & -1.6591 & 0.1929   & baseline \\ 
$H_2$ & 4 & 60 & IDBlock & 480 & 360 & -1.8520  & -0.9728 & 0.8802   & -35.64\% \\ 
 &  &  & AntiBP & 179 & 131 &      & -1.7409 & 0.1110  & 42.42\% \\ \hline
 &  &  & Vallina & 400 & 300 &  & -1.4555 & 0.3965 & baseline \\ 
$H_2$ & 4 & 100 & IDBlock & 800 & 600 & -1.8520 & -0.9681 & 0.8839 & -12.29\% \\ 
 &  &  & AntiBP & 291 & 230 &     & -1.5939 & 0.2581 & 34.91\% \\ \hline
 &  &  & Vallina & 640 & 480 &  & -1.2075 & 0.6445 & baseline \\ 
$H_2$ & 4 & 160 & IDBlock & 1280 & 960 & -1.8520 & -0.9335 & 0.9185 & -4.25\% \\ 
 &  &  & AntiBP & 449 & 332 &   & -1.3183 & 0.5336 & 17.20\% \\ \hline
\end{tabular}
\label{tab:result}
\end{center}
\end{table*}

\subsubsection{Effective Evaluation in Noisy Environment}

We optimized the expectation of molecules $H_2$ in 4 qubits circuit and $H_2 O$ in 14 qubits. As the Table ~\ref{tab:result} shows, in the noisy environment, identity block circuits perform the worst, highlighting their unsuitability for noise-prone conditions. While the identity block method mitigates barren plateaus in noiseless settings by introducing additional structure, it increases the depth of the circuit. This deeper circuit accumulates more noise, which significantly degrades performance. As a result, the identity block approach fails to achieve competitive expectation values under noisy conditions and proves to be counterproductive.

On the other hand, AntiBP circuits excel Vallina circuits in noisy environments in different depths by pruning unnecessary gates, including redundant rotation and controlled gates. This gate reduction decreases the circuit depth and limits noise accumulation, resulting in superior performance. 

These findings highlight that the identity block method, though effective in noiseless environments, is unsuitable for scenarios with noise. In contrast, architecture optimization AntiBP proves to be a robust and efficient approach for maintaining performance under noisy hardware conditions.

\begin{figure}[t]
\centerline{\includegraphics[width=0.5\textwidth]{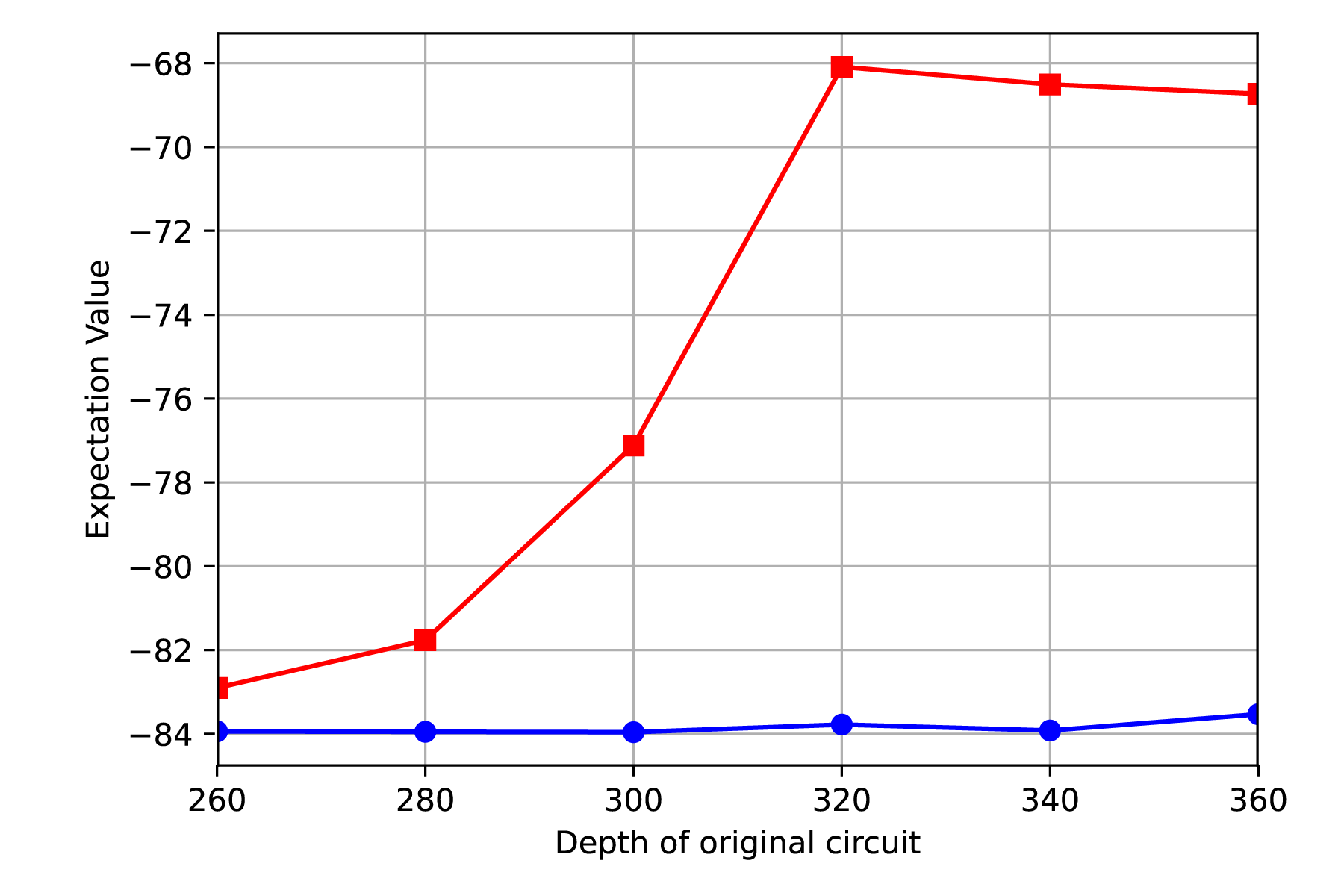}}
\caption{Performance Comparison of AntiBP and random pruning method with same pruning ratio.}     
\label{fig:ablation}
\end{figure}

\subsubsection{Ablation Study}

To further evaluate the effectiveness of our architecture search method, we conducted an ablation study by introducing a comparison experiment. In this experiment, we pruned gates (both rotation gates and controlled gates) from the baseline circuit randomly, ensuring that the total number of gates matched the count in the architecture-optimized circuit obtained through our search method. The goal was to determine whether the observed performance improvement of the architecture-optimized circuit was due merely to the reduction in gate count or to the specific gates selected during the search process.

The results, as shown in Figure~\ref{fig:ablation}, demonstrate a clear distinction between the two approaches. For the baseline circuit with random pruning, barren plateaus persisted at depths where they were previously observed in the unpruned baseline. Despite reducing the gate count, random pruning failed to address the core issue of barren plateaus because it did not systematically identify and remove gates that contribute to the problem. As a result, the randomly pruned circuit showed poor performance at deeper depths, with expectation values deteriorating significantly.

In contrast, the architecture-optimized circuit consistently achieved superior performance, even with the same number of gates. This result highlights that our architecture search method effectively identifies gates that are redundant or detrimental to convergence. By removing these gates and retaining only those critical for expressiveness, the architecture search not only reduces the gate count but also avoids barren plateaus. As a result, the architecture-optimized circuit maintained low expectation values across all depths, demonstrating its robustness and efficiency.

This ablation study confirms that the effectiveness of our method lies not just in reducing the number of gates but in intelligently identifying and pruning gates that hinder optimization. The targeted pruning ensures that the resulting circuit is both compact and capable of achieving superior convergence, even in challenging scenarios such as deep circuits.

\section{Conclusion}

In this work, we demonstrated that it is possible to automatically prune a given quantum circuit to escape barren plateaus, resulting in more efficient and robust circuits. We propose AntiBP, a framework that identifies and prunes redundant parameters and gates to address BP. By dynamically refining the circuit architecture, AntiBP removes unnecessary elements, enabling more efficient optimization and better convergence. Notably, the circuits generated by AntiBP not only escape barren plateaus in ideal, noise-free environments but also demonstrate enhanced performance when tested under realistic noisy conditions, further validating the robustness and adaptability of the approach.

\bibliographystyle{IEEEtran}


\end{document}